\documentclass[aps,preprint]{revtex4}%
\usepackage{amsfonts}
\usepackage{amsmath}
\usepackage{amssymb}
\usepackage{graphicx}%
\providecommand{\U}[1]{\protect\rule{.1in}{.1in}}
\begin{document}
\title[ ]{Concerning the Direction of the Aharonov-Bohm Deflection}
\author{Timothy H. Boyer}
\affiliation{Department of Physics, City College of the City University of New York, New
York, New York 10031, USA}

\begin{abstract}
The interaction of a solenoid with a passing charged particle can be treated
within classical or quantum physics. If charged particles pass around both
sides of a solenoid, there is an experimentally-observed Aharonov-Bohm
deflection of the double-slit particle interference pattern between charges
passing on opposite sides. \ Such a deflection can be obtained by a classical
force calculation. \ Although the magnitude of the angular deflection agrees
between the classical force calculation and the quantum topological theory,
the direction of the predicted deflection \ is opposite. \ Here we point out
the simple basis for the direction of the deflection based upon classical
electrodynamics and based upon quantum theory, and we mention analogues, both
the electrostatic deflection of the particle interference pattern and the
optical analogue of the classical calculation. \ The deflection direction
involves an experimental question which is addressed rarely if ever.

\end{abstract}
\maketitle

\section{Introduction}

When particles pass through two slits in a barrier and are recorded on a
distant screen, the intensity pattern of particle arrivals produces a
double-slit particle interference pattern (depending on the separation between
the slits) inside a single slit envelope (depending on the width of the
slits). \ If a long solenoid is placed just behind the barrier and parallel to
the slits, the magnetic field in the solenoid will produce a deflection of the
double-slit interference pattern inside the undisplaced single-slit envelope
arising for charged particles. \ Such behavior is seen in the beautiful
experiments of Moellensted and Bayh\cite{M1962}\cite{Bayh1962} involving not
slits in a barrier, but rather a biprism arrangement in an electron
microscope, where the displaced double-slit interference pattern and
undisplaced single-slit envelope are clear. \ A figure in the experimental
report\cite{Bayh1962} shows the continuous deflection of the double-beam
pattern with continuous change of the solenoid current. \ This magnetic
Aharonov-Bohm phase shift due to the solenoid\cite{AB1959} is treated in all
the recent textbooks on quantum mechanics\cite{Gq}\cite{Ball} and is also
discussed in some textbooks of classical electrodynamics.\cite{Shad}%
\cite{Garg} \ However, when physicists are asked to give the direction of the
deflection of the double-slit interference pattern, many seem to draw a blank.
\ The magnitude of the Aharonov-Bohm phase shift is presented
repeatedly.\cite{Gq}\cite{Ball} \ However, the direction of the deflection,
apparently, is another matter. \ Is the deflection in the same direction as
that taken by a charged particle experiencing a Lorentz force when passing
through the center of the solenoid, or is the deflection in the opposite
direction? \ In this note, we wish to discuss the direction of the magnetic
Aharonov-Bohm deflection from the viewpoint of classical forces and to compare
it with the quantum result which claims no forces are present. \ \ The
predictions of the two theories are in direct conflict.

\section{Electromagnetic Interactions}

According to some interpretations, the Aharonov-Bohm effect is a purely
quantum topological effect having no classical electromagnetic analogue.
\ Despite this claim, we can attempt to discuss the situation from the
familiar viewpoint of classical electromagnetic forces. \ Furthermore, the
\textit{magnitude} of the angle of deflection of the double-slit interference
pattern when viewed on a distant screen is independent of Planck's constant
and the magnitude of the deflection angle agrees completely between classical
force calculations and the quantum analysis which claims no forces are
present. \ This independence from Planck's constant and agreement in magnitude
between the classical and quantum theories also encourages the idea that
classical physics may give us some useful insights. \ Furthermore, there is an
electrostatic-force-based effect which leads to exactly the sort of particle
interference pattern shift as observed experimentally.\cite{B1987a} \ The
electrostatic-force-based deflection has been measured by Mattucci and
Pozzi.\cite{MP}

In the \textit{Feynman Lectures},\cite{F1964} it is suggested that the
magnetic Aharonov-Bohm deflection is in the same direction as that of a
particle passing through the center of the solenoid. \ Indeed, the
\textit{Lectures} propose a classical analogue involving a magnetic deflection
of the entire interference pattern. \ This suggestion turns out to be
completely wrong, as Feynman himself acknowledged.\cite{F1964} \ The
experiments\cite{Bayh1962} show clearly that only the double-slit interference
pattern is displaced, not the single-slit envelope. \ In a continuing source
of confusion, the \textit{Feynman Lectures} are still published without any
corrections, and so still contain the original errors in Vol. II, Section
15-5. \ 

\section{Classical Electromagnetic Forces Between the Charge and the Solenoid}

As far as classical theory is concerned, a good place to start on the analysis
of the deflection direction is a conventional Lorentz-force calculation. \ It
turns out that the magnetic field of a charge passing outside an
electrically-neutral solenoid places a \textit{net Lorentz force on the
solenoid}, even an infinite solenoid. This classical electromagnetic force
seems rarely (if ever) mentioned in the literature of the Aharonov-Bohm
interaction. \ \ If we consider an infinite circular solenoid of radius $R$,
surface current $\mathbf{K=}\widehat{\phi}K,$ (with $K=nI$, $n$ turns per unit
length, current $I$ in each turn) and central axis along the $z$-axis. and a
passing charge $q$ with velocity $\mathbf{v=}\widehat{y}v$ at position
$\mathbf{r=}\widehat{x}x_{0}+\widehat{y}vt$ at time $t$ with $R<<x_{0}$, then
the magnetic Lorentz force on the solenoid for a charge $q$ passing outside
the solenoid is
\begin{align}
\mathbf{F}_{onSolenoid}^{\left(  qOutside\right)  }  &  =\int_{-\infty
}^{\infty}dz\int_{0}^{2\pi}d\phi R\frac{K}{c}\left(  -\widehat{x}\sin
\phi+\widehat{y}\cos\phi\right)  \times\mathbf{B}_{q}\left(  R,\phi
,z\mathbf{,}t\right) \nonumber\\
&  =q\frac{v}{c}\frac{K}{c}R\int_{-\infty}^{\infty}dz\int_{0}^{2\pi}d\phi
\frac{\left[  -\widehat{x}\cos\phi\left(  R\cos\phi-x_{0}\right)
-\widehat{y}\sin\phi(R\cos\phi-x_{0})\right]  }{\left[  \left(  R\cos
\phi-x_{0}\right)  ^{2}+\left(  R\sin\phi-vt\right)  ^{2}+z^{2}\right]
^{3/2}}\nonumber\\
&  =\frac{2\pi qvKR^{2}}{c^{2}}\left\{  \widehat{x}\frac{x_{0}^{2}-\left(
vt\right)  ^{2}}{\left[  x_{0}^{2}+\left(  vt\right)  ^{2}\right]
}+\widehat{y}\frac{2x_{0}vt)}{\left[  x_{0}^{2}+\left(  vt\right)
^{2}\right]  ^{2}}\right\}  \label{FonS}%
\end{align}
where we have used the magnetic field of the moving charge $q$ through order
$1/c^{2}$,%
\begin{equation}
\mathbf{B}_{q}\left(  R,\phi,z,t\right)  =q\widehat{y}\frac{v}{c}\times
\frac{\left(  \widehat{x}R\cos\phi+\widehat{y}R\sin\phi+\widehat{z}%
z-\widehat{x}x_{0}-\widehat{y}vt\right)  }{\left[  \left(  R\cos\phi
-x_{0}\right)  ^{2}+\left(  R\sin\phi-vt\right)  ^{2}+z^{2}\right]  ^{3/2}},
\end{equation}
have expanded the denominator for small $R<<x_{0}$ as%
\begin{align}
&  \left[  \left(  R\cos\phi-x_{0}\right)  ^{2}+\left(  R\sin\phi-vt\right)
^{2}+z^{2}\right]  ^{-3/2}\nonumber\\
&  =\left[  x_{0}^{2}+\left(  vt\right)  ^{2}+z^{2}\right]  ^{-3/2}\left(
1+3R\frac{(x_{0}\cos\phi+vt\sin\phi}{\left[  x_{0}^{2}+\left(  vt\right)
^{2}+z^{2}\right]  }+...\right)  ,
\end{align}
and have noted the integral%
\begin{equation}
\int dz\frac{1}{\left[  z^{2}+a^{2}\right]  ^{5/2}}=\frac{3a^{2}z+2z^{3}%
}{3a^{4}\left[  z^{2}+a^{2}\right]  ^{3/2}}.
\end{equation}
In general, the force on the solenoid has a component in the same
$\widehat{y}$ direction as the velocity $\mathbf{v=}\widehat{y}v$ of the
charge $q$, and also an $\widehat{x}$ component perpendicular to the velocity.
\ The force on the solenoid in the $\widehat{y}$ (longitudinal) direction
reverses sign with $x_{0}$, corresponding to reversing sign depending on which
side the charge $q$ passes the solenoid. \ If the (positive) charge $q$ passes
on the right-hand side of the solenoid (with magnetic field upwards), the
solenoid is attracted towards the charge $q$ at both negative and positive times.

On the other hand, if the moving charge $q$ is inside the solenoid at its
center, the direction of the magnetic Lorentz force on the solenoid is
\begin{align}
\mathbf{F}_{onSolenoid}^{\left(  qInside\right)  }  &  =\int_{-\infty}%
^{\infty}dz\int_{0}^{2\pi}d\phi R\frac{K}{c}\left(  -\widehat{x}\sin
\phi+\widehat{y}\cos\phi\right)  \times\mathbf{B}_{q}\left(  R,\phi
,z\mathbf{,}t\right) \nonumber\\
&  =q\frac{v}{c}\frac{K}{c}R\int_{-\infty}^{\infty}dz\int_{0}^{2\pi}d\phi
\frac{\left[  -\widehat{x}R\cos^{2}\phi-\widehat{y}R\sin\phi\cos\phi\right]
}{\left[  R^{2}+z^{2}\right]  ^{3/2}}\nonumber\\
&  =-\widehat{x}2\pi q\frac{vK}{c^{2}},
\end{align}
which is only in the direction perpendicular to the particle velocity
$\mathbf{v}$, and in the direction opposite from the perpendicular component
of the Lorentz force acting on the solenoid due to the magnetic field of the
charge $q$ when closest to but outside the solenoid. \ \ 

\section{Forces in the Inertial Frame Where the Charge is Initially At Rest}

In general, accounts of the Aharonov-Bohm interaction between a solenoid and a
passing charge do not mention the magnetic Lorentz force of the charge on the
solenoid. \ Indeed, the experimental measurements do not treat the solenoid,
but rather measure the behavior of the passing charge. However, if a passing
charge puts a force on the solenoid, then perhaps the solenoid puts a force on
the passing charge. \ The usual claim is made that there is no force on the
passing charge due to the solenoid because the magnetic fields of an
\textit{unperturbed }infinite\textit{ }solenoid are confined to the interior
of the solenoid.\cite{B1973b} \ However, these accounts are inadequate because
they do not discuss the power delivered to the currents of the solenoid by the
emf associated with the moving charge $q$,\cite{B1973c} and the acceleration
of the charges in the solenoid producing Faraday electric fields back on the
passing charge $q$.\cite{B2023b}

In this article, we will not attempt to treat the Faraday fields of the
accelerating charges in the solenoid. \ Rather we will use Lorentz
transformations to go to the primed inertial frame in which the charge $q$ is
initially at rest while the solenoid is moving. \ Assuming at most very
low-velocity motion for the charge $q$ in this new inertial frame and lowest
order in $1/c^{2}$, we can deal merely with \textit{electrostatic} forces and
unperturbed currents in the solenoid. \ Since the charge $q$ is initially at
rest and experiences only a small force, it does not impose a significant emf
on the currents of the solenoid. \ Since the forces between the solenoid and
the charge $q$ are already of order $1/c^{2}$ (as seen in Eq. (\ref{FonS})),
the forces will be invariant between inertial frames in lowest order in powers
of $1/c^{2}$. \ 

In the primed inertial frame in which the solenoid is moving with a velocity
$-\mathbf{v}=-\widehat{y}v$, each turn of the solenoid acquires an electric
dipole moment $\mathbf{p}^{\prime}$ derived from the magnetic moment
$\mathbf{m=}\pi R^{2}I/c$ of the turn of radius $R$ carrying current
$I$,\cite{G571}\cite{J2nd}\cite{Bco}%

\begin{equation}
\mathbf{p}^{\prime}=\left(  -\widehat{y}v\right)  \times\widehat{z}I\pi
R^{2}/c=-\widehat{x}vI\pi R^{2}/c.
\end{equation}
When the turns are stacked, the entire solenoid now has a line of electric
dipoles in the primed frame. \ Just as an electric dipole can be pictured as
two opposite charges separated by a small distance $\epsilon$, the line of
electric dipoles can be pictured as two line charges $\pm\lambda$ separated by
a small distance $\epsilon$ in the $x$-direction so that the line of electric
dipoles is $-\widehat{x}\lambda\epsilon=-\widehat{x}nvI\pi R^{2}%
/c=-\widehat{x}vK\pi R^{2}/c$, where $n$ is the number of turns per unit
length of the solenoid. \ If the solenoid has a magnetic field upwards,
$\mathbf{B=}\widehat{z}4\pi nI/c$, then the $+\lambda$ is toward the left and
the $-\lambda$ is toward the right when looking in the $+\widehat{y}%
$-direction of the velocity $\mathbf{v}$. \ The direction of deflection can be
understood from the electric forces between the charge $q$ and the line of
electric dipoles. \ In the primed inertial frame, a positive charge (positron)
(passing on the right side of the solenoid in the unprimed frame) would be
attracted to the closer negative line charge $-\lambda$ and repelled by the
more distant positive line charge $+\lambda$. \ 

In the primed inertial frame, it is easy to calculate the electrostatic forces
between the line of electric dipoles in the solenoid and the charge $q$ which,
in the primed inertial frame, is initially at rest, by calculating the forces
between the charge $q$ and the line charges $\pm\lambda$, $\mathbf{F}%
_{\text{on}q}=q\mathbf{E}_{\lambda}+q\mathbf{E}_{-\lambda}$, and then
expanding in powers of $\epsilon$. \ The electric force on the electric
dipoles due to the charge $q$ is exactly that found above in Eq. (\ref{FonS}),
while the electric force on the charge $q$ due to the line charges is just the
negative. \ Thus the magnetic Lorentz force on the solenoid in one inertial
frame becomes an electric force in the primed inertial frame. \ Also,
electrostatic forces satisfy Newton's third law, so that we have obtained the
force on the charge $q$ without needing to go through the Faraday acceleration
fields which appear in the unprimed inertial frame. \ In earlier
work\cite{B1987a}, the line of electric dipoles was suggested merely as an
\textquotedblleft electrostatic analogue\textquotedblright\ of the
Aharonov-Bohm deflection. \ Here we claim that the electric forces between the
solenoid and the charge are exactly the forces found in the primed coordinate
frame. \ To order $1/c^{2}$, exactly the same forces are found in the primed
and unprimed inertial frames.\cite{G544} \ 

\section{Displacement of the Charge $q$\ }

Since through order $1/c^{2}$ the forces are the same in the primed and
unprimed inertial frames, we may apply the forces in the unprimed inertial
frame where the charge $q$ is moving with initial velocity $\mathbf{v=}%
\widehat{y}v$ when it is far from the solenoid. \ The acceleration of the
charge $q$ due to the force from the solenoid is
\begin{equation}
m\mathbf{\ddot{r}=-}\frac{2\pi qvKR^{2}}{c^{2}}\left\{  \widehat{x}\frac
{x_{0}^{2}-\left(  vt\right)  ^{2}}{\left[  x_{0}^{2}+\left(  vt\right)
^{2}\right]  ^{2}}+\widehat{y}\frac{2x_{0}vt)}{\left[  x_{0}^{2}+\left(
vt\right)  ^{2}\right]  ^{2}}\right\}  \label{mrdd}%
\end{equation}
Assuming that the force on the charge $q$ is small so that that the
acceleration of the charge is very small, the speed of the charge may be
approximated by $\mathbf{v}$ on the right-hand side of Eq. (\ref{mrdd}),
giving a small change in the velocity of the charge $q$ due to the force from
the solenoid as
\begin{align}
\Delta\mathbf{\dot{r}}_{x_{0}}\left(  t\right)   &  \mathbf{=}\mathbf{-}%
\frac{2\pi qvKR^{2}}{mc^{2}}\left\{  -\widehat{x}\int_{-\infty}^{t}dt^{\prime
}\left[  \frac{x_{0}^{2}-\left(  vt^{\prime}\right)  ^{2}}{\left[  x_{0}%
^{2}+\left(  vt^{\prime}\right)  ^{2}\right]  ^{2}}\right]  +\widehat{y}%
\int_{-\infty}^{t}dt^{\prime}\frac{x_{0}vt^{\prime}}{\left[  x_{0}^{2}+\left(
vt^{\prime}\right)  ^{2}\right]  ^{2}}\right\} \nonumber\\
&  =\mathbf{-}\frac{2\pi qvKR^{2}}{mc^{2}}\left\{  \widehat{x}\frac
{vt}{v\left[  x_{0}^{2}+\left(  vt\right)  ^{2}\right]  }-\widehat{y}%
\frac{x_{0}}{v\left[  x_{0}^{2}+\left(  vt\right)  ^{2}\right]  }\right\}  ,
\label{Drd}%
\end{align}
which can be checked by direct differentiatition. For the $\widehat{y}$
(longitudinal) direction parallel to $\mathbf{v}$, the velocity change of the
charge in Eq. (\ref{Drd}) is of only one sign depending on the sign of $x_{0}%
$, and vanishes at early and late times, $\Delta\mathbf{\dot{r}}\rightarrow0$
for $t\rightarrow\pm\infty$. \ However, in the $\widehat{x}$ direction
perpendicular to $\mathbf{v}$, the charge $q$ acquires a small velocity which
is in opposite directions as the charge $q$ approaches the solenoid at
negative times or leaves the solenoid at positive times. \ 

Finally, using the same approximations and integrating again with respect to
time from $t\rightarrow-\infty$ to $t\rightarrow\infty$, we obtain the net
relative displacement of the charge $q$ due to the force from the solenoid, as
a relative lag in the longitudinal direction parallel to $\mathbf{v}$,
\begin{equation}
\Delta\mathbf{r}_{x_{0}}\mathbf{=\pm}\widehat{y}\frac{2\pi^{2}qKR^{2}}%
{vmc^{2}}, \label{Drpd}%
\end{equation}
with the $\pm$ sign depending upon the sign of $x_{0}$ corresponding to the
side on which the passing took place. \ In the approximation which treats the
change in the charge velocity as very small, the net relative displacement
$\Delta\mathbf{r}$ for particles located on opposite sides of the solenoid is
thus entirely in the longitudinal direction and is doubled compared to Eq.
(\ref{Drpd})
\begin{equation}
\Delta\mathbf{r}=\Delta\mathbf{r}_{x_{0}}-\Delta\mathbf{r}_{-x_{0}}%
=\pm\widehat{y}\frac{4\pi^{2}qKR^{2}}{vmc^{2}}=\pm\frac{q\Phi}{vmc},
\end{equation}
where $\Phi=\pi R^{2}B=\pi R^{2}\left(  4\pi K/c\right)  $ is the magnetic
flux in the solenoid. \ Noting our use of gaussian units here, the magnitude
of the relative displacement is exactly that suggested in earlier in
references \cite{B1973c}) and \cite{B1987a} where SI units were used. \ In the
unprimed inertial frame in which the solenoid is at rest, these changes will
be superimposed upon the motion of the charge $q$ with velocity $\mathbf{v}$. \ 

For a positive charge $q$ passing the solenoid on the right-hand side where
$x_{0}$ is positive and with the solenoid magnetic field upwards, there is
always an attractive force in the $\widehat{y}$ direction of motion on the
charge $q$, accelerating the charge at negative times and then slowing the
charge at positive times as $t$ changes from negative to positive values.
\ The reversal of the longitudinal force on opposite sides of the solenoid (as
$x_{0}$ changes sign) leads to a relative longitudinal displacement $\Delta y$
between positive charges passing on opposite side of the solenoid where the
displacement $x_{0}$ has opposite signs. \ This longitudinal displacement of
the particles in the direction of motion leads to an angular deflection
$\theta~$for a wavefront connecting the charges where, for small displacements
$\Delta y$ and slits separated by a distance $d$,
\begin{equation}
\theta=\Delta y/d=\pm q\Phi/(pcd), \label{theta}%
\end{equation}
where $\Phi$ is the magnetic flux in the solenoid in gaussian units and $p=mv$
is the momentum of the passing charge $q$. \ \ The magnitude of this classical
angular deflection is independent of Planck's constant $\hbar$, and is exactly
that of the quantum treatment. \ The direction of deflection would be to the
left in the direction of the particle's motion as described above.\ 

\section{The Optical Analogy}

We have already noted that the experimentally observed deflection of the
double-beam \textit{particle} interference pattern for the \textit{magnetic}
Aharonov-Bohm deflection takes the same form as the deflection based upon
classical \textit{electrostatic} forces.\cite{B1987a} \ The shift in the
\textit{particle} interference pattern is analogous to an the \textit{optical}
effect where a piece of glass is placed behind only one slit in the
barrier.\cite{B1987a} \ Then the light which passes through the piece of glass
will be slowed, and we will find an angular deflection $\theta$ in the
double-slit interference pattern toward the side with the retarding glass,
which deflection depends upon the relative lag $\Delta y$ in the direction of
propagation between the phases of the light waves passing through the two
slits separated by distance $d$, giving an angular deflection for small angles
$\theta=\Delta y/d$. $\ $On the other hand, the single-slit envelope is
undisplaced since it depends upon wave interference across each single slit.
\ The angle of the optical \textit{deflection} depends upon the length of the
optical path through the glass and is completely independent of the wavelength
of the radiation, just as the angular Aharonov-Bohm \textit{deflection} is
independent of Planck's constant and of any associated de Broglie wavelength.
\ In principle, the angular deflection could be converted into a relative lag
between wave packets passing through the two slits if the optical beam were
chopped into packets of finite length. \ 

The classical electromagnetic force analysis suggests that the magnetic
Aharonov-Bohm deflection is like both the electrostatic force-based deflection
and the optical analogy. \ Thus we expect the interaction of the charged
particle and the solenoid to give a deflection of the interference pattern
through an angle $\theta=\pm q\Phi/\left(  cpd\right)  $; the direction of the
deflection corresponds to a classical lag arising in the longitudinal
direction for the charge $q$, caused by a force opposite to the longitudinal
component of the magnetic Lorentz force of the charged particle on the
solenoid. \ The direction of deflection for the double-slit interference
pattern in the classical force-based magnetic Aharonov-Bohm interaction for
charges passing outside a solenoid is in the opposite direction from those
passing through the center of the solenoid which experience an unambiguous
Lorentz-force deflection. \ 

\section{Quantum Deflection Direction}

The quantum magnetic Aharonov-Bohm deflection is based upon the classical
Hamiltonian for a particle in the presence of a time-invariant magnetic field
$\mathbf{B=\nabla\times A}$,
\begin{equation}
H=\frac{1}{2m}\left[  \mathbf{p}_{\text{canonical}}\mathbf{-}\frac{q}%
{c}\mathbf{A(r)}\right]  ^{2}=\frac{1}{2m}\mathbf{p}_{\text{mechanical}}^{2},
\end{equation}
where the canonical momentum is connected to the mechanical momentum by
$\mathbf{p}_{\text{canonical}}=\mathbf{p}_{\text{mechanical}}+\left(
q/c\right)  \mathbf{A(r)}$. \ The time dependent Schroedinger equation follows
as%
\begin{equation}
i\hbar\frac{\partial\psi}{\partial t}=\frac{1}{2m}\left[  \frac{\hbar}%
{i}\nabla\mathbf{-}\frac{q}{c}\mathbf{A(r)}\right]  ^{2}\psi.
\end{equation}
The solution of the Schroedinger equation is
\begin{equation}
\psi\left(  \mathbf{r,}t\right)  =C\exp\left\{  \frac{i}{\hbar}\left[
\mathbf{p\cdot}\int^{\mathbf{r}}d\mathbf{r}^{\prime}+\frac{q}{c}%
\int^{\mathbf{r}}d\mathbf{r}^{\prime}\cdot\mathbf{A}\left(  \mathbf{r}\right)
-\mathcal{E}t\right]  \right\}  , \label{psirt}%
\end{equation}
where here the momentum $\mathbf{p}$ is a constant \textit{mechanical}
momentum associated with a constant particle speed which is connected to the
constant mechanical energy $\mathcal{E=}\mathbf{p}^{2}/\left(  2m\right)  $.
\ The phase change associated with this wave function is given in the quantum
texts\cite{Gq}\cite{Ball} and leads to a magnitude of the angle of the
double-slit particle interference pattern which is identical with the
magnitude of the classical force-based deflection given in Eq. (\ref{theta}).

The direction of the quantum deflection depends upon the variation of the path
length $\int^{\mathbf{r}}d\mathbf{r}^{\prime}$ multiplying the constant
mechanical momentum $\mathbf{p}$ in Eq. (\ref{psirt}). \ If the particle
passes the solenoid on the side such that $\left(  q/c\right)  \int%
^{\mathbf{r}}d\mathbf{r}^{\prime}\cdot\mathbf{A}\left(  \mathbf{r}\right)  $
is positive, then the path length must be smaller to compensate the phase.
\ Thus for a positive charge $q$ passing on the right-hand side of a solenoid
with the magnetic field $\mathbf{B}$ upwards, the quantum deflection would be
to the right, in the same direction as the magnetic Lorentz force deflection
of a particle passing through the center of the solenoid.

\section{Direct Conflict Between Classical And Quantum Results}

The quantum direction is exactly the opposite direction from that predicted by
the classical force-based calculation given previously. \ The quantum
calculation also involves no interaction between the passing charge and the
solenoid. \ In sharp contrast, the classical electrodynamic force-based
analysis depends upon an electromagnetic interaction between the passing
charge and the solenoid, in agreement with the interaction for the classical
electrostatic force-base deflection\cite{B1987a} measured by Matteucci and
Pozzi,\cite{MP} and in agreement with the interaction between the light and
the piece of glass in the optical analogy. \ However, the quantum result seems
to claim that the charged particle passes the solenoid with no interaction, no
change in particle velocity, and a purely topological effect is involved.

Although the author has looked for the observed deflection direction in the
beautiful experimental measurements, he has been unable to find the needed
information. \ 

\section{Acknowledgement}

I wish to thank Professor V. Parameswaran Nair for helpful discussions related
to the direction of the Aharonov-Bohm deflection. \

\bigskip April 14, 2023 \ ABdeflection4.tex

\end{document}